\documentclass[amsmath,11pt]{article}

\usepackage{amsfonts,amsmath,amssymb,graphics,epsfig}

\usepackage{appendix}

\date{\empty}

\begin{document}

\author{Fani Dosopoulou${}^{1,2}$ and Christos G. Tsagas${}^{3,4}$\\ {\small ${}^1$Department of Physics and Astronomy, Northwestern University}\\ {\small 2145 Sheridan Road, Evanston, IL, 6028, USA}\\ {\small ${}^2$Department of Physics, University of Crete, Heraklion 71003, Greece}\\ {\small ${}^3$Section of Astronomy, Astrophysics and Mechanics, Department of Physics}\\ {\small Aristotle University of Thessaloniki, Thessaloniki 54124, Greece}\\ {\small ${}^4$Theoretical Astrophysics, Eberhard Karls University of T\"ubingen, T\"ubingen 72076, Germany}}

\title{\bf Vorticity survival in magnetised Friedmann universes}

\maketitle

\begin{abstract}
We use a general relativistic approach to investigate the effects of weak cosmological magnetic fields on linear rotational perturbations during the radiation and dust epochs of the universe. This includes ordinary kinematic vorticity, as well as vortex-like inhomogeneities in the density distribution of the cosmic medium. Our study confirms that magnetic fields can source both types of perturbations and that their presence helps cosmic rotation to survive longer. In agreement with previous Newtonian studies, we find that during the dust era vorticity decays slower than in non-magnetised cosmologies. The relativistic nature of the treatment means that we can also investigate the epoch prior to equipartition. There, the magnetic effect is more pronounced, since it helps both of the aforementioned rotational distortions to maintain constant magnitude throughout the radiation era. Overall, magnetised universes not only generate vorticity but also provide a much better environment for the survival of rotational perturbations, compared to their magnetic-free counterparts.
\end{abstract}

\section{Introduction}\label{sI}
Rotation is a common phenomenon in the universe, as most astrophysical bodies rotate. Over the years, this has lead a number of authors to raise the question of global rotation, whether or not, in other words, the whole cosmos rotates as well (see~\cite{H} for a representative though incomplete list). After all general relativity allows for rotating spacetimes, with Godel's solution being perhaps the most celebrated and inspirational example~\cite{G,OS}. Although we should not expect a definite answer to the question of cosmic rotation any time soon, there has been speculation as to whether certain anisotropic features of the Cosmic Microwave Background (CMB) could be explained by small amounts of large-scale vorticity~\cite{Jetal}. This brings to the fore the next question, which is finding physical mechanisms that could generate rotation on cosmological scales. Perturbation theory can provide some answers. It has been known, in particular, that there is no vorticity generation at the linear perturbative level, if the cosmic medium remains ideal.\footnote{This does not generally apply to ``tilted'' cosmological models, where the observers have a peculiar velocity (a ``tilt'' angle) relative to the fundamental reference frame (e.g.~see~\cite{HdPIC}).} In that case, one needs to go to the nonlinear stage in order to induce rotational distortions~\cite{MB}. Viscosity, on the other hand, can act as a source of vorticity at the linear level and the same is also true for magnetic fields. Viscous effects can also change the standard evolution of rotational distortions in perturbed Friedmann-Robertson-Walker (FRW) cosmologies~\cite{R}-\cite{DdSTB}. This happens because ``imperfections'' in the equation of state of the various matter fields that fill the universe lead to forces which can source vorticity and affect its linear evolution as well. Neutrino vorticities, in particular, were found to remain constant during the radiation era on superhorizon scales~\cite{L1}.

Magnetic fields are also quite ubiquitous in the universe and their presence has been repeatedly verified on all but the largest (cosmological) scales~\cite{K}. As with viscosity, it is the generic anisotropy of the $B$-field that generates vorticity~\cite{W}. More specifically, to linear order, it is the tension component of the Lorentz force which triggers rotational perturbations. It has been shown that such distortions can survive on small scales (below the Silk-damping threshold) in the photon-baryon plasma~\cite{JKO}. This could lead to potentially observable signatures in the CMB, a possibility that has attracted considerable interest and investigation (e.g.~see~\cite{MKK}). Here we will focus on the magnetic implications for pre-existing vorticity rather than the role of the field as a source of rotational perturbations.  Employing a Newtonian analysis, it was shown that magnetic fields can help vorticity to survive longer, by slowing down the standard decay-rate of linear rotational distortions associated with perfect-fluid Friedmann-Robertson-Walker (FRW) cosmologies~\cite{DdSTB}. An analogous magnetic effect on linear vector (vortex-like) density inhomogeneities has also been noted in relativistic, dust-dominated Friedmann models~\cite{BMT}. Overall, it appears that magnetised cosmologies could contain more residual rotation than their non-magnetised counterparts. The aim of the present work is to investigate further this possibility, by extending the previous studies into the fully relativistic regime.

We begin with an introduction to the kinematics of rotating observers and a brief reference to basic aspects of relativistic magnetohydrodynamic (MHD) theory. Our next step is to consider a non-magnetised FRW universe filled with a highly conductive perfect fluid, which implies that we will be working within the ideal MHD approximation. Perturbing this background, we allow for the presence of a weak magnetic field and then examine how it affects the linear rotational behaviour of our model. After a brief discussion of non-magnetised vorticity perturbations, primarily for comparison reasons, we consider the field's implications for both ordinary kinematic vorticity and for vortex-like inhomogeneities in the density distribution of the medium. As expected, we confirm that magnetic fields generally act as sources of rotational distortions and also affect their evolution. More specifically, we provide for the first time (to the best of our knowledge) analytical solutions monitoring the linear evolution of magnetised rotational perturbations during the radiation and the dust eras. These solutions, which are fully relativistic, show that linear vorticity perturbations and density vortices decay slower in magnetised than in magnetic-free cosmologies. During the radiation era, in particular, both of the aforementioned types of rotational distortions remain constant, instead of decaying at a rate inversely proportional to the dimensions of the universe. After equilibrium, the field's presence also slows down the ``standard'' (non-magnetised) depletion rate of rotational perturbations, giving the latter a better chance of surviving. Consequently, magnetised universes are expected to rotate faster and longer than their magnetic-free counterparts. Quantitatively speaking, we find that the former models have approximately twenty orders of magnitude larger residual vorticity than the latter, assuming the same initial conditions. This means that, in principle at least, magnetised cosmologies can start off with considerably smaller amounts of rotation and still sustain appreciable levels of it today.

\section{Kinematics of rotating observers}\label{sKROs}
In accord with the 1+3 covariant formulation of general relativity (see~\cite{TCM} for a recent review), the kinematics of a family of observers is determined by a set of irreducible variables that describe the relative motion of their worldlines. The aforementioned quantities obey three pairs of propagation and constraint equations, all of which follow from the Ricci identities.

\subsection{The irreducible variables}\label{ssIVs}
Consider a 4-dimensional spacetime with a Lorentzian metric $g_{ab}$ of signature ($-,+,+,+$) and introduce a family of (fundamental) observers moving with 4-velocity $u_a$. The latter is tangent to the observers' timelike worldlines, namely  $u^a={\rm d}x^a/{\rm d}\tau$, where $x^a=x^a(\tau)$ and $\tau$ is the associated proper time, so that $u_au^a=-1$. The $u_a$-field defines the time direction, while the symmetric tensor $h_{ab}=g_{ab}+u_au_b$ projects orthogonal to $u_a$ and into the observers' instantaneous 3-dimensional rest space. Then, overdots indicate (proper) time differentiation and ${\rm D}_a= h_a{}^b\nabla_b$ defines the 3-D covariant derivative operator, with $\nabla_a$ representing the 4-D covariant derivative (e.g.~$\dot{u}_a=u^b\nabla_bu_a$ and ${\rm D}_bu_a= h_b{}^dh_a{}^c\nabla_du_c$ -- see Eq.~(\ref{Nbua}) below).\footnote{By construction $h_{ab}u^b=0$, $h_{ac}h^c{}_b=h_{ab}$, $h_a{}^a=3$ and ${\rm D}_ch_{ab}=0$. Note that, when there is no rotation, the projector $h_{ab}$ also acts as the metric tensor of the spatial hypersurfaces orthogonal to the $u_a$-field.}

Local variations in the observers' motion are monitored by the gradient of their 4-velocity field, which is decomposed into the  irreducible kinematic variables as follows
\begin{eqnarray}
\nabla_bu_a&=& {\rm D}_bu_a- A_au_b \nonumber\\ &=&{1\over3}\,\Theta h_{ab}+ \sigma_{ab}+ \omega_{ab}- A_au_b\,.  \label{Nbua}
\end{eqnarray}
In the above $\Theta=\nabla^au_a={\rm D}^au_a$ is the volume expansion/contraction scalar, $\sigma_{ab}={\rm D}_{\langle b}u_{a\rangle}$ is the symmetric and trace-free shear tensor, $\omega_{ab}={\rm D}_{[b}u_{a]}$ is the antisymmetric vorticity tensor and $A_a=\dot{u}_a$ is the 4-acceleration vector.\footnote{Round brackets denote symmetrisation and square ones antisymmetrisation. Angled brackets indicate the symmetric and trace-free part of an orthogonally projected (spacelike) second-rank tensor (e.g.~$\sigma_{ab}={\rm D}_{\langle b} u_{a\rangle}= {\rm D}_{(b}u_{a)}-({\rm D}^cu_c/3)h_{ab}$), or the spatial component of a vector (e.g.~$\dot{\omega}_{\langle a\rangle}=h_a{}^b\dot{\omega}_b$ -- see Eq.~(\ref{dotomega1}) below).} The last three of these variables are spacelike by construction, namely they satisfy the constraints  $A_au^a=0=\sigma_{ab}u^b= \omega_{ab}u^b$.

The volume scalar tracks the average relative motion between the worldlines of neighbouring observers. In particular, positive values for $\Theta$ indicate volume expansion and negative ones contraction. This scalar is also used to introduce a representative length scale ($a$), defined by $\dot{a}/a=\Theta/3$. Changes in the shape of the worldline congruence, under constant volume, are encoded in the shear, while the vorticity monitors their rotational behaviour. Note that the antisymmetry of the vorticity tensor ensures that it has only three independent componets, which means we can replace it with the vorticity vector $\omega_a= \varepsilon_{abc}\omega^{bc}/2$. The latter is also spacelike (i.e.~$\omega_au^a=0$) and defines the rotational axis of the relative motion.\footnote{By definition, $\varepsilon_{abc}= \eta_{abcd}u^d$ is the totally antisymmetric 3-dimensional Levi-Civita tensor, with $\eta_{abcd}$ being its 4-D counterpart. Also, $\dot{\varepsilon}_{abc}= 3u_{[a}\varepsilon_{bc]d}A^d$, ${\rm D}_d\varepsilon_{abc}=0$ and $\varepsilon_{abc}\varepsilon^{def}= 3!h_{[a}{}^dh_b{}^eh_{c]}{}^f$ by construction~\cite{TCM}.} Finally, the 4-acceleration reflects the presence of non-gravitational forces and vanishes when the observers move under gravity alone, in which case their worldlines are timelike geodesics.

\subsection{Propagation equations and constraints}\label{ssPECs}
The irreducible kinematic variables of the previous section obey a set of three propagation formulae that are supplemented by an equal number of constrains. All are derived after applying the Ricci identities to the observers' 4-velocity vector, namely by means of $2\nabla_{[a}\nabla_{b]}u_c=R_{abcd}u^d$, with $R_{abcd}$ representing the Riemann curvature tensor. In practice, this means splitting the Ricci identities into their timelike and spacelike components and then isolating the trace, the symmetric trace-fee and the antisymmetric parts of the resulting relations~\cite{TCM}. The propagation equations are\footnote{We use geometrised units, with $\kappa=8\pi G=1=c$, throughout this manuscript.}
\begin{equation}
\dot{\Theta}= -{1\over3}\,\Theta^2- {1\over2}\,\left(\rho+3p\right)- 2\left(\sigma^2-\omega^2\right)+ {\rm D}^aA_a+ A_aA^a\,,  \label{Ray1}
\end{equation}
\begin{equation}
\dot{\sigma}_{\langle ab\rangle}= -{2\over3}\,\Theta\sigma_{ab}- \sigma_{c\langle a}\sigma^c{}_{b\rangle}- \omega_{\langle a}\omega_{b\rangle}+ {\rm D}_{\langle a}A_{b\rangle}+ A_{\langle a}A_{b\rangle}- E_{ab}+ {1\over2}\,\pi_{ab}  \label{dotsigma1}
\end{equation}
and
\begin{equation}
\dot{\omega}_{\langle a\rangle}= -{2\over3}\,\Theta\omega_a- {1\over2}\,{\rm curl}A_a+ \sigma_{ab}\omega^b\,,  \label{dotomega1}
\end{equation}
where $\rho$ is the energy density of the matter, $p$ is its isotropic pressure and $\pi_{ab}$ its anisotropic (viscous) counterpart (with $\pi_{ab}=\pi_{ba}$, $\pi^a{}_a=0=\pi_{ab}u^b$). Also, $E_{ab}$ is the so-called electric Weyl tensor, which is primarily associated with the tidal part of the (long-range) gravitational field. Finally, $\sigma^2=\sigma_{ab}\sigma^{ab}/2$ and $\omega^2=\omega_{ab}\omega^{ab}/2=\omega_a\omega^a$ define the magnitudes of the shear and the vorticity respectively. The kinematic constrains, on the other, hand read
\begin{equation}
{\rm D}^a\omega_a= A^a\omega_a\,,  \label{vortcon1}
\end{equation}
\begin{equation}
H_{ab}= {\rm curl}\sigma_{ab}+ {\rm D}_{\langle a}\omega_{b\rangle}+ 2A_{\langle a}\omega_{b\rangle}  \label{mWeylcon1}
\end{equation}
and
\begin{equation}
{\rm D}^b\sigma_{ab}= {2\over3}\,{\rm D}_a\Theta+ {\rm curl}\omega_a+ 2\varepsilon_{abc}A^b\omega^c- q_a\,.  \label{shearcon1}
\end{equation}
with $H_{ab}$ representing the magnetic component of the Weyl field and $q_a$ the energy flux vector of the matter (so that $q_au^a=0$). Note that both Weyl tensors are symmetric, trace-free and spacelike by construction  (i.e.~$E_{ab}=E_{ba}$, $H_{ab}=H_{ba}$, $E^a{}_a=0=H^a{}_a$ and $E_{ab}u^b=0=H_{ab}u^b$). Also, ${\rm curl}v_a=\varepsilon_{abc}{\rm D}^bv^c$ for any spacelike vector $v_a$ and ${\rm curl}v_{ab}=\varepsilon_{cd\langle a}{\rm D}^c v_{b\rangle}{}^d$ for any spacelike, symmetric and traceless tensor $v_{ab}$.

Expression (\ref{dotomega1}) is the key equation for our purposes, since it monitors the rotational behaviour of neighbouring worldlines. This formula ensures that there is no vorticity generation unless non-gravitational forces are included into the system (i.e.~$\omega_a=0\rightarrow\dot{\omega}_a=0$, unless $A_a\neq0$). Alternatively, one could say that irrotational timelike geodesics remain so. The same formula can also be used to track changes in the rotational axis of the motion, namely effects like precession and nutation. As we have mentioned at the beginning, forces that generate rotation can come from a variety of sources, including viscosity, non-barotropicity and magnetic fields. These agents can affect the evolution of a rotating fluid as well. Here, we will focus on magnetic fields and consider their implications for the generation, the evolution and the survival of rotational distortions in the context of cosmology.

\subsection{Aspects of rotating spaces}\label{ssARSs}
Most of the available studies assume non-rotating worldline congruences, which in technical terms ensures that the associated 4-velocity field is hypersurface orthogonal. This in turn means that there are integrable 3-dimensional surfaces forming the common rest spaces of all the fundamental observers at a given instant of time. Rotation changes all these. The observers' worldlines are no longer hypersurface orthogonal and their instantaneous rest-spaces do not mesh together to form a single 3-dimensional surface. As a result, even the spatial gradients of scalars do not commute. Instead, we have
\begin{equation}
{\rm D}_{[a}{\rm D}_{b]}\phi= -\omega_{ab}\dot{\phi}\,,  \label{3Ricci1}
\end{equation}
for any given scalar $\phi$ and
\begin{equation}
2{\rm D}_{[a}{\rm D}_{b]}v_c= -2\omega_{ab}h_c{}^d\dot{v}_d+ \mathcal{R}_{dcba}v^d\,,  \label{3Ricci2}
\end{equation}
for a spacelike vector $v_a$. These are the so-called 3-Ricci identities, with $\mathcal{R}_{abcd}$ being the 3-dimensional Riemann tensor. The latter monitors the intrinsic geometry of the observers rest spaces and is related to its spacetime counterpart ($R_{abcd}$) by means of
\begin{equation}
\mathcal{R}_{abcd}= h_a{}^eh_b{}^fh_c{}^qh_d{}^sR_{efqs}- {\rm D}_cu_a{\rm D}_du_b+ {\rm D}_du_a{\rm D}_cu_b\,.  \label{3Riemann}
\end{equation}
Starting from the 3-Riemann tensor, one can define the 3-Ricci tensor and the associated 3-Ricii scalar as $\mathcal{R}_{ab}= h^{cd}\mathcal{R}_{acbd}$ and $\mathcal{R}= h^{ab}\mathcal{R}_{ab}$ respectively (see \S~1.3.5 of~\cite{TCM} and also Appendix A.3 there for more details). We should also note that the orthogonally projected Ricci identities, especially Eq.~(\ref{3Ricci1}), play a key role in the evolution of rotating spacetimes (e.g.~see \S~\ref{ssLVPs} below).

\section{Conservation laws}\label{sCLs}
To proceed, we need to specify our medium and derive the corresponding conservation laws. We will assume a single, highly conductive perfect fluid. The high electrical conductivity implies that we will be working within the ideal MHD approximation. Technically speaking, this means that the electric fields vanish in the observers' rest frame and the currents keep the magnetic component of the Maxwell field ``frozen'' into the matter.

\subsection{Magnetic energy conservation}\label{ssMEC}
The vanishing of the electric fields is guaranteed by Ohm's law. The latter takes the covariant form $\mathcal{J}_a=\varsigma E_a$, with $\mathcal{J}_a$ representing the spatial current density (with $\mathcal{J}_au^a=0$) and and $\varsigma$ the electrical conductivity of the medium~\cite{Gr}. At the ideal MHD limit, where $\varsigma\rightarrow\infty$, Ohm's law ensures that there are no electric fields in the observers' frame. As a result, Maxwell's equations reduce to a set of one propagation and three constraint equations, respectively given by
\begin{equation}
\dot{B}_{\langle a\rangle}= -{2\over3}\,\Theta B_a + (\sigma_{ab}+\varepsilon_{abc}\omega^c)B^b  \label{M1}
\end{equation}
and
\begin{equation}
\mathcal{J}_a= {\rm curl}B_a+\varepsilon_{abc}A^bB^c\,, \hspace{10mm} 2\omega_aB^a= \rho_e\,, \hspace{10mm} {\rm D}^aB_a=0\,,  \label{M234}
\end{equation}
where $\rho_e$ is the electric charge density. The former of the above ensures that the magnetic forcelines connect the same matter particles at all times, which implies that the field is frozen into the highly conductive medium. Also, contracting Eq.~(\ref{M1}) along the $B_a$ vector and taking into account that $B^2=B_aB^a$, we arrive at
\begin{equation}
\left(B^2\right)^{\cdot}= -{4\over3}\,\Theta B^2- 2\sigma_{ab}\Pi^{ab}\,,  \label{dotB2}
\end{equation}
which is the conservation law of the magnetic energy density. Note that $\rho_B=B^2/2$ is the magnetic energy density, $p_B=B^2/6$ represents the isotropic pressure of the field and $\Pi_{ab}=-B_{\langle a}B_{b\rangle}$ is the magnetic anisotropic stress tensor (with $\Pi_{ab}=\Pi_{ba}$ and $\Pi^a{}_a=0=\Pi_{ab}u^b$)~\cite{BMT}.

\subsection{Matter energy and momentum conservation}\label{ssMEMC}
Our perfect fluid assumption means that both the energy-flux vector and the anisotropic stress tensor of the matter vanish identically (i.e.~$q_a=0=\pi_{ab}$). Under these conditions, the energy and momentum conservation laws of our highly conductive magnetised medium are
\begin{equation}
\dot{\rho}= -\Theta(\rho+p)  \label{ce1}
\end{equation}
and
\begin{equation}
\left[\left(\rho+p+{2\over3}\,B^2\right)h_{ab}+\Pi_{ab}\right]A^b= -{\rm D}_ap- \varepsilon_{abc}B^b{\rm curl}B^c\,,  \label{NSe1}
\end{equation}
respectively.\footnote{On the left-hand side of Eq.~(\ref{NSe1}), we see how the energy density, the isotropic pressure and the anisotropic stresses of the magnetic field also contribute to the total effective inertial ``mass'' of the system.} The former of the above expressions is the relativistic continuity equation and the latter can be seen as the magnetised version of the Navier-Stokes formula. Note that there are no magnetic terms in the right-hand side of the continuity equation. This, together with the absence of explicit matter terms in Eq.~(\ref{dotB2}), ensures that (at the ideal MHD limit) the magnetic and the matter energy densities are separately conserved. In contrast, both sources contribute to the Navier-Stokes equation, which governs the conservation of the momentum density.

Expression (\ref{NSe1}) is the second key equation for our purposes. Here, the main magnetic input comes from the Lorentz force, which conveniently splits into a pressure and a tension stress as
\begin{equation}
\varepsilon_{abc}B^b{\rm curl}B^c= {1\over2}\,{\rm D}_aB^2- B^b{\rm D}_bB_a\,.  \label{Lorentz}
\end{equation}
The first term on the right-hand side is due to the (positive) magnetic pressure and the second comes from the field's tension, namely from the negative pressure the $B$-field exerts along its own direction. Recall that the former tends to push the magnetic forcelines apart, while the latter reflects their elasticity and tendency to remain ``straight''. As we will see in the following sections, the magnetic effects on vorticity come mainly from the field's tension properties.

\section{Rotating almost-FRW universes}\label{sRA-FRWUs}
Before we start looking into the magnetic effects on rotating, almost-FRW universes, we should briefly discuss the rotational behaviour of non-magnetised cosmological models. In either case, our starting point is a magnetic-free Friemdmannian background that contains a single barotropic perfect fluid. In the magnetised case, the cosmic medium will also be highly conductive.

\subsection{The background cosmology}\label{ssBC}
The symmetry (isotropy and homogeneity) of the FRW spacetimes ensures that the only nonzero variables are scalars that depend solely on time. Hence, the only physical quantities allowed in a Friedmann model are the energy density and the isotropic pressure of the matter, with $p=p(\rho)$ for barotropic media, the Hubble parameter, defined as $H=\Theta/3=\dot{a}/a$, and the background 3-Ricci scalar $\mathcal{R}=6K/a^2$, where $K=0,\pm1$ is the 3-curvature index. In the absence of a cosmological constant, this background evolves in line with the zero-order continuity equation,
\begin{equation}
\dot{\rho}=- 3H(1+w)\rho\,,  \label{FRWce}
\end{equation}
supplemented by Friedmann's formulae
\begin{equation}
H^2= {1\over3}\,\rho- {K\over a^2} \hspace{10mm} {\rm and} \hspace{10mm} \dot{H}= -H^2- {1\over6}\,(1+3w)\rho\,.  \label{FRWs1}
\end{equation}
Here, $w=p/\rho$ is the barotropic index that determines the nature of the matter. A related thermodynamic variable is the adiabatic sound speed, the square of which is defined as $c_s^2=\dot{p}/\dot{\rho}$. Note that $c_s^2$ coincides with the barotropic index, when the latter is time invariant (i.e.~$c_s^2=w$ when $\dot{w}=0$ and vice versa -- see \S~3.2.1 in~\cite{TCM}).

Once the geometry of the 3-dimensional hypersurfaces and the nature of the matter component have been specified, the above system closes and can be solved analytically. In the case of Euclidean spatial geometry and radiation, for example, we may set $K=0$ and $w=1/3$. Then, Eqs.~(\ref{FRWce}) and (\ref{FRWs1}) lead to the familiar solution $a\propto t^{1/2}$, $H=1/2t$ and $\rho=3/4t^2$. When dealing with pressureless dust, on the other hand, we find that $a\propto t^{2/3}$, $H=2/3t$ and $\rho=4/3t^2$.

\subsection{Linear vorticity perturbations}\label{ssLVPs}
The equations given in \S~\ref{sKROs} and \S~\ref{sCLs} earlier are fully nonlinear and apply to any spacetime, provided that matter is described by a single fluid. Let us momentarily ignore the magnetic presence and linearise these formulae around an FRW background. When doing so, quantities with nonzero background value will be assigned zero perturbative order, while those that vanish there will be treated as first-order (gauge-invariant) variables. Also note that the temporal and spatial derivatives of perturbed variables retain their original perturbative order. Finally, when linearising, terms of perturbative order higher than the first are dropped from our equations, all of which reduces Eq.~(\ref{dotomega1}) to
\begin{equation}
\dot{\omega}_a= -2H\omega_a- {1\over2}\,{\rm curl}A_a\,.  \label{ldotomega1}
\end{equation}
Similarly, the non-magnetised version of the Navier-Stokes equation (see expression (\ref{NSe1}) in \S~\ref{ssMEMC} earlier) linearises to Euler's formula
\begin{equation}
\rho(1+w)A_a= -{\rm D}_ap\,.  \label{lmdc1}
\end{equation}

Combining the above and keeping in mind that the 3-D gradients of scalars do not commute in rotating spaces (see Eq.~(\ref{3Ricci1}) in \S~\ref{ssARSs}), provides the linear evolution equation of the vorticity vector within an almost-FRW universe
\begin{equation}
\dot{\omega}_a= -2\left(1-{3\over2}\,c_s^2\right)H\omega_a\,.  \label{ldotomega2}
\end{equation}
The above, which holds for all three types of background spatial curvature, shows that pressure gradients cannot generate vorticity at the linear level. If the fluid is already rotating, the aforementioned gradients also leave the rotational axis unaffected, but they generally effect the rate of rotation and thus the residual amount of vorticity. When there is no pressure, in particular, vorticity simply decays with the universal expansion as $\omega\propto a^{-2}$ on all scales. Once pressure has been introduced, however, this decay rate slows down. In the case of radiation, for example, the solution of Eq.~(\ref{ldotomega2}) gives $\omega\propto a^{-1}$. Further increase in the pressure of the rotating medium, can even reverse the decay. More specifically, for barotropic matter ``stiffer'' than $c_s^2=2/3$, vorticity increases with the expansion. This ``reversal'' is a purely general relativistic effect. It reflects the absence of a spatial hypersurface of simultaneity, common to all rotating observers, and results from the non-commutativity of the 3-D gradients of scalars seen in Eq.~(\ref{3Ricci1}).

Applying the above to the post-inflationary evolution of an almost-FRW universe we have $\omega_{eq}=\omega_0(a_0/a_{eq})= \omega_0(T_{eq}/T_0)$ at equilibrium and $\omega_*=\omega_{eq}(a_{eq}/a_*)^2=\omega_{eq}(T_*/T_{eq})^2$ at present.~\footnote{The zero suffix denotes a given initial time, which here will always assume that it coincides with the beginning of the radiation epoch. The $*$-suffix, on the other hand, corresponds to the present.} Note that $T_0$, $T_{eq}$ and $T_*$ are respectively the temperatures at the beginning of the radiation era, at the time of matter-radiation equality and today. Also, recall that $T\propto a^{-1}$ throughout the lifetime of the universe. Finally, setting $T_*\simeq10^{-13}$~GeV, $T_{eq}\simeq10^4T_*\simeq10^{-9}$~GeV $T_0\simeq10^{10}$~GeV, which is close to the typical reheating temperature, we arrive at
\begin{equation}
\omega_*= \left({T_*^2\over T_0T_{eq}}\right)\omega_0\simeq 10^{-27}\omega_0\,,  \label{omega*1}
\end{equation}
for the residual value of a given vorticity mode at present. Therefore, the current amount of cosmic rotation (in a non-magnetised universe) is about 27 orders of magnitude below its value at the onset of the radiation era.

From the observational point of view, a more practical variable is the dimensionless ratio $\omega/H$, giving the amount of universal rotation relative to the average expansion of the background universe. During the radiation epoch $\omega\propto a^{-1}$ and $H\propto a^{-2}$, which means that $\omega/H\propto a$ before equipartition. After equilibrium $\omega\propto a^{-2}$ and $H\propto a^{-3/2}$, ensuring that $\omega/H\propto a^{-1/2}$ throughout the dust era. These evolution laws immediately translate into $(\omega/H)_{eq}=(\omega/H)_0(a_{eq}/a_0)$ and $(\omega/H)_*=(\omega/H)_{eq}(a_*/a_{eq})^{-1/2}$, which combine to give
\begin{equation}
\left({\omega\over H}\right)_*= \left({\omega\over H}\right)_0 \left({T_0T_*^{1/2}\over T_{eq}^{3/2}}\right)\simeq 10^{17}\left({\omega\over H}\right)_0\,,  \label{nmomega/H}
\end{equation}
for the same temperature values used in Eq.~(\ref{omega*1}) earlier. The above provides the relative rotation of the universe at present, in terms of its value at the beginning of the radiation era. Current observations constrain the ratio $\omega/H$ to very small values. Following~\cite{Jetal}, in particular, we may set $(\omega/H)_*\sim 10^{-10}$ (higher upper limits for this ratio have also been quoted in the literature -- e.g.~see~\cite{B}). Then, expression (\ref{nmomega/H}) implies that $(\omega/H)_0\sim10^{-27}$ initially. In the following sections, we will see how a magnetic presence (even a weak one) can change these results.

\section{Magnetised rotating almost-FRW 
universes}\label{sMRA-FRWUs}
Magnetic fields seem to be everywhere in the universe, since their presence has been repeatedly verified in galaxies, in galaxy clusters and also in high-redshift young proto-galactic clouds. Moreover, recently, there have been claims for the first ever magnetic detection in intergalactic voids. All these have made the idea of primordial magnetism particularly appealing.

\subsection{Linearising around a Friedmann 
background}\label{ssLAFRWB}
Let us consider an almost-FRW universe permeated by a weak large-scale magnetic field. The weakness of the latter means that it will be treated as a perturbation upon the aforementioned Friedmannian background. We will therefore always impose the constraint $B^2/\rho\ll1$, to guarantee that the magnetic contribution to the total energy-momentum tensor is well below that of the dominant matter component. This means that $B^2$ will be treated as a first-order perturbation, which in turn implies that the magnetic vector ($B_a$) and its gradients ($\dot{B}_a$ and ${\rm D}_bB_a$) are half-order distortions.\footnote{Given that $B^2=B_aB^a$ is a perturbation of order one, the $B_a$-field is of order half. Also, since ${\rm D}_aB^2=2B^b{\rm D}_aB_b$ is a first-order distortion, the spatial gradient ${\rm D}_aB_b$ has order 1/2. Therefore, all the magnetic terms in our linear equations are of perturbative order one, which guarantees the consistency of our linearisation scheme.} Then, the key nonlinear expressions (see Eqs.~(\ref{dotomega1}) and (\ref{NSe1}) in \S~\ref{ssPECs} and \S~\ref{ssMEMC} respectively) read
\begin{equation}
\dot{\omega}_a= -2H\omega_a- {1\over2}\,{\rm curl}A_a\,,  \label{lmdotomega1}
\end{equation}
and
\begin{equation}
\rho(1+w)A_a= -{\rm D}_ap- \varepsilon_{abc}B^b{\rm curl}B^c\,.  \label{lmdc2}
\end{equation}
Also, to lowest perturbative order, the magnetic field evolves according to the set (see relations (\ref{M1}) and (\ref{M234}) in \S~\ref{ssMEC})
\begin{equation}
\dot{B}_a= -2HB_a \hspace{10mm} {\rm and} \hspace{10mm} {\rm D}^aB_a= 0\,.  \label{ldotB}
\end{equation}

The set (\ref{lmdotomega1})-(\ref{ldotB}) governs the rotational behaviour of a weakly magnetised almost-FRW universe, filled with a highly conductive perfect fluid. To account for the magnetic effects, we need to combine Eqs.~(\ref{lmdotomega1}), (\ref{lmdc2}) and employ a rather lengthy calculation, the details of which are in Appendix~A at the end of this paper. The result is the linear vorticity propagation formula
\begin{equation}
\dot{\omega}_a= -2H\left(1-{3\over2}\,c_s^2\right)\omega_a- {1\over2\rho(1+w)}\left(B^b{\rm D}_b{\rm curl}B_a-{\rm curl}B^b{\rm D}_{(b}B_{a)}\right)\,.  \label{lmdotomega2}
\end{equation}
According to the above, $B$-fields can act as sources of rotational distortions at the linear perturbative level. In fact, it is the elasticity of the magnetic forcelines that triggers these perturbations, since both of the source-term on the right-hand side of (\ref{lmdotomega2}) come from the tension component of the Lorentz force (see Appendix~A below). Note that the aforementioned magnetic terms are of perturbative order one. which reconfirms the  consistency of from our linearisation.

\subsection{Incorporating the magnetic effect}\label{ssIME}
The relative importance of the two magnetic terms on the right-hand side of Eq.~(\ref{lmdotomega2}) depends on the degree of the inhomogeneity of the perturbed spacetime. By construction, the second term tends to dominate in highly inhomogeneous environments and for this reason its contribution was neglected in previous studies. Here, we will go one step further and account for all the linear magnetic effects. Despite this additional ``complication'', one can still solve Eq.~(\ref{lmdotomega2}) analytically, by taking its time derivative and then eliminating the magnetic term from the right-hand side. To achieve this we also need the auxiliary linear expressions
\begin{equation}
(B^b{\rm D}_b{\rm curl}B_a)^{\cdot}= -6HB^b{\rm D}_b{\rm curl}B_a  \label{1orderaux1}
\end{equation}
and
\begin{equation}
({\rm curl}B^b{\rm D}_{(b}B_{a)})^{\cdot}= -6H{\rm curl}B^b{\rm D}_{(b}B_{a)}\,.  \label{1orderaux2}
\end{equation}
Note that both of the above result from the linear commutation law $({\rm }D_bv_a)^{\cdot}={\rm D}_b\dot{v}_a-H{\rm D}_bv_a$, which holds for any first-order spacelike vector $v_a$ and on all FRW backgrounds (e.g.~see Eq.~(A.32) in Appendix A.3 of~\cite{TCM}).

Taking the time derivative of (\ref{lmdotomega2}), using the the background relations (\ref{FRWce}), (\ref{FRWs1}) and the linear commutation laws (\ref{1orderaux1}), (\ref{1orderaux2}), we arrive at
\begin{eqnarray}
\ddot{\omega}_a&=& -2\left(1-{3\over2}\,c_s^2\right)H\dot{\omega}_a+ 2\left(1-{3\over2}\,c_s^2\right) \left[1+{1\over2}\,(1+3w)\Omega\right]H^2\omega_a \nonumber\\ &&+{3(1-w)H\over2(1+w)\rho}\left(B^b{\rm D}_b{\rm curl}B_a-{\rm curl}B^b{\rm D}_{(b}B_{a)}\right)\,,  \label{lmddotomega1}
\end{eqnarray}
with $\Omega=\rho/3H^2$ representing the density parameter of the universe. Finally, by going back to Eq.~(\ref{lmdotomega2}), we can express the magnetic term at the end of the above with respect to vorticity. Then, the differential equation (\ref{lmddotomega1}) recasts into
\begin{equation}
\ddot{\omega}_a= -5\left(1-{6\over5}\,w\right)H\dot{\omega}_a- 4\left(1-{3\over2}\,w\right) \left[1-{3\over2}\,w-{1\over4}\,(1+3w)\Omega\right]H^2\omega_a\,,  \label{lmddotomega2}
\end{equation}
which no longer contains explicit magnetic terms. Note that in the process we have set $\dot{w}=0$, which in turn implies that $c_s^2=w$ (see also \S~\ref{ssBC} earlier). For all practical purposes, this assumption does not affect the generality of expression (\ref{lmddotomega2}). Recall that the equation of state of the matter is expected to remain invariant during prolonged periods in the lifetime of our universe (throughout the radiation and dust ears for example).

The above formula monitors the linear evolution of rotational perturbations on a weakly magnetised highly conductive almost-FRW background. This is a new fully relativistic differential equation that incorporates all the linear magnetic effects, including those that were bypassed previously. Expression (\ref{lmddotomega2}) is surprisingly simple and as a result of this it can be solved analytically, at least when the background spatial geometry is Euclidean.

\subsection{Evolution in the radiation era}\label{ssERE}
Let us consider the radiation epoch first. Setting $w=1/3$ in the right-hand side of (\ref{lmddotomega2}) and introducing the harmonic splitting $\omega_a=\sum_n\omega_{(n)}\mathcal{Q}_a^{(n)}$, with ${\rm D}_a\omega_{(n)}=0=\dot{\mathcal{Q}}_a^{(n)}$ and ${\rm D}^2\mathcal{Q}_a^{(n)}=-(n/a)^2\mathcal{Q}_a^{(n)}$, expression (\ref{lmddotomega2}) reduces to
\begin{equation}
\ddot{\omega}_{(n)}= -3H\dot{\omega}_{(n)}- \left(1-\Omega\right)H^2\omega_{(n)}\,.  \label{lmrddotomega1}
\end{equation}
The latter holds on all scales, provided the ideal-MHD approximation applies there.\footnote{On a spatially flat background, the eigenvalue ($n$) of the vorticity mode coincides with its comoving wavenumber, while $\lambda_n=a/n$ (with $n>0$) is the associated physical wavelength.} Current observations indicate that $|1-\Omega|\lesssim10^{-3}$ at present, supporting a nearly flat universe. Consequently, on an FRW background with Euclidean spatial geometry, where $\Omega=1$, $a\propto t^{1/2}$ and $H=1/2t$, the above differential equation takes the form
\begin{equation}
\ddot{\omega}_{(n)}= -{3\over2t}\,\dot{\omega}_{(n)}  \label{lmrddotomega2}
\end{equation}
and accepts the power-law solution
\begin{equation}
\omega_{(n)}= \mathcal{C}_1+ \mathcal{C}_2t^{-1/2}= \mathcal{C}_1+ \mathcal{C}_3a^{-1}\,,  \label{lmromega1}
\end{equation}
on all scales (recall that $a\propto t^{1/2}$ before equipartition). The magnetic presence has therefore added a new constant mode to the vorticity evolution law, leaving the original decaying mode unaffected (compare to the magnetic-free solutions given in \S~\ref{ssLVPs} earlier). Finally, after evaluating the integration constants in Eq.~(\ref{lmromega1}), we arrive at
\begin{equation}
\omega= \omega_0+ {\dot{\omega}_0\over H_0}\left(1-{a_0\over a}\right)\,,  \label{lmromega2}
\end{equation}
where the zero suffix indicates the onset of the radiation era and we have dropped the mode-index ($n$). Hence, in the magnetic presence, linear vorticity perturbations no longer decay as $\omega\propto a^{-1}$, but tend to constant. In particular, following solution (\ref{lmromega2}), we find that
\begin{equation}
\omega_{eq}\simeq \omega_0+ {\dot{\omega}_0\over H_0}\,,  \label{momegaeq}
\end{equation}
at the time of matter-radiation equality. Before closing, we should point out that viscosity can have an analogous effect on rotational distortions. Neutrino vortices also remain constant during the radiation era on superhorizon scales~\cite{L1}. Here, the responsible agent is the magnetic field and the affected region extends to all scales where the ideal-MHD limit applies.

\subsection{Evolution in the dust era}\label{ssEDE}
Moving to the subsequent epoch of dust domination, while maintaining the spatial flatness of the FRW background spacetime, we have $w=0$, $\Omega=1$, $a\propto t^{2/3}$ and $H=2/3t$. Proceeding as before, we find that in this new environment the differential equation (\ref{lmddotomega2}) reduces to
\begin{equation}
\ddot{\omega}_{(n)}= -{10\over3t}\,\dot{\omega}_{(n)}- {4\over3t^2}\,\omega_{(n)}\,.  \label{lmdddotomega1}
\end{equation}
Recalling that $a\propto t^{2/3}$ after equipartition, the above solves to give
\begin{equation}
\omega_{(n)}= C_1t^{-1}+ C_2t^{-4/3}= C_3a^{-3/2}+ C_4a^{-2}\,,  \label{lmdomega1}
\end{equation}
on all scales where the ideal-MHD limit applies.. Again, the $B$-field has added a mode to the linear solution, without affecting the ``standard'' one (see \S~\ref{ssLVPs} for comparison). Similarly to the radiation case, the new (magnetically induced) mode decays slower than its original (magnetic-free) counterpart. Finally, after evaluating the integration constants, the above recasts into
\begin{equation}
\omega\simeq 2\left(2\omega_{eq}+{\dot{\omega}_{eq}\over H_{eq}}\right)\left({a_{eq}\over a}\right)^{3/2}\,,  \label{lmdomega2}
\end{equation}
at late times (i.e.~for $a\gg a_{eq}$). Comparing the above to the magnetic-free case, we notice that the $B$-field slows down the decay-rate of vorticity perturbations from $\omega\propto a^{-2}$ to $\omega\propto a^{-3/2}$. This result is in full agreement with that obtained through the Newtonian analysis of~\cite{DdSTB}, ensuring that the magnetic presence helps vorticity to survive during the dust epoch as well. Overall, our analysis suggests that magnetised cosmologies should contain more residual rotation than their magnetic-free counterparts. Next, we will attempt to quantify this statement.

\subsection{The residual cosmic vorticity}
Keeping only the dominant mode in the right-hand side of solution (\ref{lmdomega1}) immediately gives $\omega_*\simeq \omega_{eq}(a_{eq}/a_*)^{3/2}\simeq10^{-6}\omega_{eq}$ for the residual vorticity today, having set $1+z_{eq}\simeq10^4$ for simplicity (recent observations suggest that $1+z_{eq}\simeq 3.5\times10^3$). Recall that the $*\,$-\,suffix corresponds to the present and $\omega_{eq}=\omega_0+\dot{\omega}_0/H_0$ is the vorticity at equilibrium (see Eq.~(\ref{momegaeq}) above). A more robust calculation makes little difference, giving
\begin{equation}
\omega_*\simeq 4\times10^{-6}
\left(\omega_0+{\dot{\omega}_0\over H_0}\right)
\end{equation}
for the current value of the vorticity. To simplify the calculation, let us assume that $\dot{\omega}_0/H_0=2\dot{\omega}_0t_0\sim \omega_0$.  Then, according to the above result, vorticity drops by approximately six orders of magnitude since the beginning of the radiation era. This should be compared to the value of $\omega_*\simeq10^{-27}\omega_0$, obtained in magnetic-free universes (see \S~\ref{ssLVPs} earlier). Consequently, the residual vorticity in weakly magnetised almost-FRW cosmologies should be approximately twenty one orders of magnitude larger than in the corresponding magnetic-free models.

The same conclusions and numerical results can be obtained by looking at the dimensionless $\omega/H$ ratio. In accord with the analysis given in \S~\ref{ssERE} and \S~\ref{ssEDE} previously, we have $\omega/H\propto a^2$ throughout the radiation epoch and $\omega/H=$~constant during the subsequent dust era. Putting these evolution laws together, while setting $T_0\simeq10^{10}$~GeV and $T_{eq}\simeq10^{-9}$~GeV as before, gives
\begin{equation}
\left({\omega\over H}\right)_*= \left({\omega\over H}\right)_0\left({T_0\over T_{eq}}\right)^2\simeq 10^{38}\left({\omega\over H}\right)_0\,,  \label{momega/H}
\end{equation}
at present. Assuming that the left-hand side of the above and that of its non-magnetised analogue (see Eq.~(\ref{nmomega/H}) in \S~\ref{ssLVPs}) are equal, we find that the ratio $(\omega/H)_0$ is approximately 21 orders of magnitude lower in the magnetised case.\footnote{Setting $(\omega/H)_*\sim10^{-10}$, as in \S~\ref{ssLVPs} for the non-magnetised case, we find that $(\omega/H)_0\sim10^{-48}$ (recall that $T_0\simeq10^{10}$~Gev and $T_{eq}\simeq10^{-9}$~GeV). In the absence of the $B$-field, the corresponding value was $(\omega/H)_0\sim10^{-27}$.} Consequently, magnetised cosmologies can start off with much lower amounts of vorticity, than their magnetic-free counterparts, and still sustain the same residual rotation today. Whether this is enough to produce astrophysically interesting levels of vorticity at present, depends on the initial value of the latter (i.e.~on $\omega_0$, or equivalently on $\omega_0/H_0$), which is treated here as a free (though always very small) parameter.

\section{Linear magnetised density vortices}\label{sLMDVs}
In addition to kinematic vorticity, magnetic fields can also source and affect density vortices. The latter are vector-type inhomogeneities, which describe rotational distortions in the matter distribution and are geometrically related to ordinary vorticity perturbations.

\subsection{Isolating the density vortices}\label{ssIDVs}
Density inhomogeneities come in three different forms: scalar, vector and tensor. The former describe overdensities/underdensities in the matter distribution and are commonly referred to as density perturbations. Inhomogeneities of vector nature are related to density vortices, while (trace-free) tensor perturbations monitor changes in the shape of the inhomogeneity, under constant volume. In what follows, we will consider the second type of these density inhomogeneities and investigate their evolution in the presence of a cosmological magnetic field.

Spatial variations in the density distribution of the matter between two neighbouring (fundamental) observers are described by the dimensionless gradient $\Delta_a=(a/\rho){\rm D}_a\rho$~\cite{TCM}. This variable contains collective information about all of the aforementioned three types of density inhomogeneities. One can decoded this information by taking the comoving gradient $\Delta_{ab}=a{\rm D}_b\Delta_a$ and then introducing the irreducible decomposition
\begin{equation}
\Delta_{ab}= {1\over3}\,\Delta h_{ab}+ W_{ab}+ \Sigma_{ab}\,.  \label{Deltaab}
\end{equation}
The scalar $\Delta=\Delta^a{}_a$ describes overdensities/underdensities in the matter distribution, when it takes positive/negative values respectively. The antisymmetric tensor $W_{ab}=\Delta_{[ab]}$ tracks density vortices and the symmetric and trace-free tensor $\Sigma_{ab}=\Delta_{\langle ab\rangle}$ is associated to shape distortions. Clearly, both $W_{ab}$ and $\Sigma_{ab}$ are spacelike by construction (i.e.~$W_{ab}u^b=0=\Sigma_{ab}u^b$). Also, in analogy with the vorticity tensor, the antisymmetry of the 3-dimensional tensor $W_{ab}$ implies that it can be replaced by the spacelike vector $W_a=\varepsilon_{abc}W^{bc}/2$. Next, we will consider the linear evolution of $W_a$ within weakly magnetised, almost-FRW universe.

\subsection{Linear magnetised density vortices}\label{ssLMDVs}
The relation between $W_a$ and the vorticity vector $\omega_a$ is more than a simple analogy, reflecting the way these two variables have been defined. In fact, the geometrical framework of general relativity guarantees that these vectors are directly connected to each other. This connection comes through the 3-Ricci identities and in particular via Eq.~(\ref{3Ricci1}), which applied to the density of the matter gives
\begin{equation}
{\rm D}_{[b}{\rm D}_{a]}\rho= \dot{\rho}\omega_{ab}\,.  \label{3Riccirho}
\end{equation}
Starting from the above, using the background continuity equation (see expression (\ref{FRWce}) in \S~\ref{ssBC}) and recalling that $W_{ab}=\Delta_{[ab]}=a^2{\rm D}_{[b}{\rm D}_{a]}\rho$ to linear order, we arrive at
\begin{equation}
W_a= -3(1+w)a^2H\omega_a\,.  \label{W-omega}
\end{equation}
This is a purely general relativistic (geometrical) result, connecting vortices in the density distribution of the matter to vorticity proper. Hence, once the background scale factor and Hubble parameter have been decided, the linear evolution of $W_a$ is essentially dictated by that of $\omega_a$.

During the radiation era we have $w=1/3$ and $a\propto t^{1/2}$, which means that $3(1+w)a^2H=4a_0^2H_0=$~constant. Therefore, before equipartition, density vortices evolve exactly as kinematic vorticity perturbations, namely
\begin{equation}
W_{(n)}= \mathcal{C}_5+ \mathcal{C}_6t^{-1/2}\,,  \label{lmrW1}
\end{equation}
for all $n>0$. In other words, rotational distortions in the density distribution of the matter remain constant throughout the radiation epoch. After equilibrium $w=0$ and $a\propto a^{2/3}$, ensuring that $a^2H=3a_0^2H_0(t/t_0)^{1/3}$. The latter combines with solution (\ref{lmdomega1}) to give
\begin{equation}
W_{(n)}= \mathcal{C}_7t^{-2/3}+ \mathcal{C}_8t^{-1}\,,  \label{lmdW1}
\end{equation}
ensuring that throughout the dust era the dominant $W$-mode decays as $W\propto t^{-2/3}$ on all scales. The same result has also been obtained through an alternative approach (see~\S~10.3 in~\cite{BMT} and references therein). Note that density vortices in non-magnetised cosmologies decay as $W\propto t^{-1/2}$ during radiation and $W\propto t^{-1}$ for dust (e.g.~see \S~3.2.5 in \cite{TCM}). Therefore, as with vorticity proper, the magnetic presence slows down the decay-rate of rotational (i.e.~vector-type) density inhomogeneities, on all scales where the ideal-MHD approximation holds.

\section{Discussion}\label{sD}
Current observations are consistent with small amounts of universal rotation, which means that, if the universe rotates, it does so very slowly. This is also in agreement with the inflationary scenario, where the exponential expansion is expected to essentially eliminate any traces of primordial vorticity. Nevertheless, most (if not all) astrophysical structures rotate, which raises the question whether their rotation is of cosmological origin, or a relatively recent addition due to local physical processes. It is not possible to generate vorticity, at the linear perturbative level, if the cosmic medium is an ideal fluid. Viscous matter fields, on the other hand, can trigger linear rotational distortions. Magnetic fields are sources of (effective) viscosity and have long been known to generate rotational perturbations. The responsible agent is the Lorentz force and more specifically its tension component. Viscosity and magnetism can also affect the linear evolution of cosmic vorticity by reducing its standard depletion rate. Newtonian studies of the magnetic effects on rotation revealed that the $B$-field slows down considerably the linear decay of vorticity perturbations after equilibrium.

The present study revisits and extends the Newtonian results using a fully relativistic approach. Assuming a weakly magnetised almost-FRW universe, we have looked into the magnetic effects on the evolution of linear rotational perturbations. These distortions, which include ordinary kinematic vorticity as well as vortex-like density inhomogeneities, could have been triggered by the $B$-field itself, or by another independent agent (or by both). Here, we have not looked into mechanisms of vorticity generation, but into the effects of the $B$-field on linear rotational perturbations. We have derived, for the first time (to the best of our knowledge) the general relativistic equations that describe the linear evolution of rotational distortions in the presence of a large-scale magnetic field. Overcoming technical problems phased by previous analogous studies, we were able to include all the magnetic effects and still solve the resulting differential equations analytically. With some adjustment, depending on the problem at hand, our equations and results could be of use in a range of cosmological applications. To revisit, for example, the magnetic effects on vector modes in the CMB spectrum.

Qualitatively speaking, our main result is that even a weak magnetic presence can help rotational distortions to survive longer than in non-magnetised models. During the radiation era, in particular, we found that the $B$-field keeps linear vorticity perturbations constant. After equilibrium, the magnetic presence slows down the standard (non-magnetised) decay-rate of these distortions. Within the geometrical framework of general relativity, kinematic vorticity and rotational density inhomogeneities are directly related. Exploiting this (linear) relation, we found that the magnetic effects on vortex-like density perturbations are exactly analogous with those on vorticity proper. Overall, magnetised cosmologies appear to rotate faster and longer than magnetic-free models. Alternatively, one could say that a magnetised universe can start off with considerably smaller amounts of initial vorticity, relative to its non-magnetised counterpart, and still sustain the same rotation levels today. Our analysis has quantified this initial difference to approximately twenty orders of magnitude.

We have arrived at the aforementioned theoretical conclusions and numerical results, by employing a linear perturbative study and by adopting the ideal-MHD approximation. The latter holds in highly conductive media and requires the presence of electric currents, which eliminate the electric fields and freeze their magnetic counterparts into the matter. These currents are generated after inflation, as the conductivity of the universe starts growing, by local physical processes and for this reason their coherence scale can never exceed that of the horizon. In other words, causality confines the electric currents and therefore their domain of influence within the particle horizon, which after inflation coincides with the Hubble radius. Beyond the Hubble scale there can be no coherent electric currents, which means that the ideal-MHD limit does not apply there. On very small scales, on the other hand, the nonlinear effects start becoming important and the linear approximation is expected to break down. All these mean that our analysis, and the conclusions derived from it, have an optimum range of scales, which roughly varies between the size of a proto-galaxy and that of the observable universe.\\

\noindent\textbf{Acknowledgments:} The authors would like to thank Theodoros Tomaras for helpful discussions and comments. F.D. acknowledges support from a  ``Maria Michail Manasaki'' Bequest Fellowship by the Department of Physics at the University of Crete and from a University Fellowship by the Department of Physics and Astronomy at Northwestern University. C.G.T wishes to thank the Section of Theoretical Astrophysics at Eberhard Karls University in T\"ubingen, where this work was completed, for their hospitality and support.\\\\

\appendix

\noindent\textbf{\Large Appendix}

\section{The vorticity propagation formula}\label{sA}
Splitting the Lorentz force into its pressure and tension parts, taking the curl of the Navier-Stokes equation (see Eqs.~(\ref{Lorentz}) and (\ref{lmdc2}) respectively) and keeping up to linear order terms, leads to the intermediate relation
\begin{equation}
\rho(1+w){\rm curl}A_a= -\varepsilon_{abc}{\rm D}^{[b}{\rm D}^{c]}p- {1\over2}\,\varepsilon_{abc}{\rm D}^{[b}{\rm D}^{c]}B^2+ \varepsilon_{ab}{}^c{\rm D}^bB^d{\rm D}_dB_c+ \varepsilon_{ab}{}^cB^d{\rm D}^b{\rm D}_dB_c\,.  \label{app1}
\end{equation}
Note that, of the three magnetic terms, the first comes from the field's (positive) pressure and the last two are due to its tension. Next, we will individually evaluate all the terms on the right-hand side of the above. Using the commutation law for the spatial gradients of scalars (see Eq.~(\ref{3Ricci1}) in \S~\ref{ssARSs} earlier), the first term linearises to
\begin{equation}
\varepsilon_{abc}{\rm D}^{[b}{\rm D}^{c]}p= 6Hc_s^2\rho(1+w)\omega_a\,,  \label{app2}
\end{equation}
while the second is zero to first order. The vanishing of the magnetic pressure term in (\ref{app1}), means that the field's tension is the sole player at the linear perturbative level. Note that, when deriving the above, we have also used the background energy conservation law and the definition $c_s^2=\dot{p}/\dot{\rho}$, of the adiabatic sound speed (see Eqs.~(\ref{FRWce}) and (\ref{ldotomega2}) in \S~\ref{ssBC} and \S~\ref{ssLVPs}) respectively). Splitting the gradients ${\rm D}_bB_a$ into their symmetric and skew parts, the third term on the right-hand side of (\ref{app1}) successively gives
\begin{eqnarray}
\varepsilon_{ab}{}^c{\rm D}^bB^d{\rm D}_dB_c&=& \varepsilon_{ab}{}^c{\rm D}^{(b}B^{d)}{\rm D}_{(d}B_{c)}+ \varepsilon_{ab}{}^c{\rm D}^{(b}B^{d)}{\rm D}_{[d}B_{c]} \nonumber\\ &&+\varepsilon_{ab}{}^c{\rm D}^{[b}B^{d]}{\rm D}_{(d}B_{c)}+ \varepsilon_{ab}{}^c{\rm D}^{[b}B^{d]}{\rm D}_{[d}B_{c]} \nonumber\\ &=&-{\rm curl}B^b{\rm D}_{(b}B_{a)}\,,  \label{app3}
\end{eqnarray}
since $\varepsilon_{ab}{}^c{\rm D}^{(b}B^{d)}{\rm D}_{(d}B_{c)}=0= \varepsilon_{ab}{}^c{\rm D}^{[b}B^{d]}{\rm D}_{[d}B_{c]}$ and $\varepsilon_{ab}{}^c{\rm D}^{(b}B^{d)}{\rm D}_{[d}B_{c]}= \varepsilon_{ab}{}^c{\rm D}^{[b}B^{d]}{\rm D}_{(d}B_{c)}= -{\rm curl}B^b{\rm D}_{(b}B_{a)}/2$. Finally, on using the 3-Ricci identities (see relation (\ref{3Ricci2}) in \S~\ref{ssARSs}), we can rewrite the last term on the right-hand side of (\ref{app1}) as
\begin{eqnarray}
\varepsilon_{ab}{}^cB^d{\rm D}^b{\rm D}_dB_c&=& B^b{\rm D}_b{\rm curl}B_a- \varepsilon_a{}^{bc}\mathcal{R}_{dbfc}B^dB^f \nonumber\\ &=&B^b{\rm D}_b{\rm curl}B_a\,,  \label{app4}
\end{eqnarray}
given that $\varepsilon_a{}^{bc}\mathcal{R}_{dbfc}B^dB^f=0$ at the linear level (irrespective of the background spatial curvature). Our last step is to combine the auxiliary formulae (\ref{app2})-(\ref{app4}) and recast Eq.~(\ref{app1}) into
\begin{equation}
{\rm curl}A_a= -6Hc_s^2\omega_a+ {1\over\rho(1+w)}\left(B^b{\rm D}_b{\rm curl}B_a-{\rm curl}B^b{\rm D}_{(b}B_{a)}\right)\,.  \label{app5}
\end{equation}
Substituting this result into expression (\ref{lmdotomega1}), one immediately arrives at the linear propagation equation (\ref{lmdotomega2}), which monitors the evolution of the vorticity vector.

\end{document}